\documentclass[aps,pre,showpacs,showkeys,preprint]{revtex4}

\usepackage{graphicx}
\usepackage{latexsym}
\usepackage{times}

\begin{document}

\title{Power-law distributions from additive preferential redistributions}

\author{Suhan Ree}
\email{suhan@kongju.ac.kr}
\affiliation{Department of Industrial Information,
Kongju National University, Yesan-Up, Yesan-Gun, 
Chungnam, 340-702, South Korea}

\date{\today}

\begin{abstract}
We introduce a non-growth model that generates the power-law 
distribution with the Zipf exponent. 
There are $N$ elements, each of which is characterized by
a quantity, and
at each time step these quantities are redistributed through 
binary random interactions with a simple additive preferential rule,
while the sum of quantities is conserved.
The situation described by this model is similar to those of 
closed $N$-particle systems when conservative two-body collisions
are only allowed.
We obtain stationary distributions of these quantities
both analytically and numerically while varying parameters of the model,
and find that the model exhibits the scaling behavior
for some parameter ranges.
Unlike well-known growth models, this alternative mechanism generates
the power-law distribution when the growth is not expected 
and the dynamics of the system is based on interactions between elements. 
This model can be applied to some examples such as personal wealths,
city sizes, and the generation of scale-free networks when only rewiring 
is allowed.
\end{abstract}



\maketitle

\setcounter{totalnumber}{1}

\section{Introduction}
Power-law distributions have been observed in diverse fields for more than a century \cite{newman}. 
Some well-known examples
exhibiting  `scaling' behavior are city sizes \cite{zipf,gabaix,zanette}, 
word frequencies \cite{simon}, sizes of business  firms \cite{gibrat},
personal incomes \cite{champernowne,pareto}, personal wealths \cite{ispolatov,chatterjee,xie}, 
sizes of web sites \cite{huberman}, numbers  of links of web pages \cite{albert}, 
connections of routers in the Internet \cite{faloutsos}, species in  genera \cite{yule}, 
interactions of proteins \cite{jeong}, citations of scientific papers \cite{redner}
and so  on, covering many research fields such as biology, 
economics, sociology,  engineering, and physics. 
Many generative models have been introduced so far to explain 
this ubiquitous phenomenon \cite{mitzenmacher}, 
and most of them use simple mechanisms 
that give rise to the power-law distributions.
One group of models uses stochastic multiplicative processes \cite{champernowne,levy,gabaix}, 
and another group uses preferential growing mechanisms \cite{yule,simon,barabasi}.   
These models have their root in the Gibrat's law of proportional growth \cite{gibrat},
and are based on two assumptions: the growth of the system and non-interaction between elements. 
There are also non-growth models in which the main mechanism is the interaction between 
randomly chosen elements, resulting in the multiplicative changes of values \cite{ispolatov,abraham,fujihara,chatterjee}. 
Systems showing the scaling behavior consist of $N$ elements ($N$ may vary with time), 
while each element $i$ ($1\leq i \leq N$)
is represented by the quantity $k_i$, and the  probability of an element having the value $k$, 
$P(k)$, has the form $k^{-\gamma}$ for a given  range of $k$.

Here we introduce a non-growth model exhibiting the power-law distribution 
with the Zipf exponent ($\gamma=2$), in which quantities of elements are redistributed through binary
random interactions with a simple additive preferential rule. 
The model assumes that
$N$ and the sum of all $k_i$'s are conserved, and that, when two elements $i$ and $j$ are chosen
randomly at a given time, $k_i$ and $k_j$ will be changed additively while preserving $k_i+k_j$.
This model can be a mechanism that explains scaling behavior of many socio-economical systems, 
especially when the growth is not expected and interactions between elements 
are vital to their dynamics. 
Moreover, this model can be extended to generate
scale-free networks through rewiring only, because the rewiring process by changing an end point
of a link changes degrees of two nodes additively while preserving the sum of degrees of all nodes. 

In this paper, the model and its stationary distributions are investigated 
both numerically and analytically.
In Sec.\ II, the model is described in detail. 
In Sec.\ III, the master equation is obtained.
Stationary distributions are found numerically first, and then analytically solved.
And the condition for the power-law distributions in
the parameter space is also found using both numerical and analytic methods.
In Sec.\ IV, three possible applications of this model are discussed.
Finally, in Sec.\ V, we summarize our results.

\section{Model}
Let us introduce our stochastic model in detail.
The model  assumes 
that $k_i$'s are non-negative integers, and we define   
$\alpha$ as the average quantity  per element,
\begin{equation}
	\alpha  \equiv \mathop \Sigma \limits_{k = 0}^\infty  kP(k) = \langle k \rangle.
\label{alpha}
\end{equation}
At each time step $T$, two elements, $i$  and $j$, 
are randomly chosen, and the element $i$ gives one unit of the quantity to the element $j$ 
with the {\em exchange probability} $R$; hence their quantities are  changed additively, 
$k_i\rightarrow k_i-1$ and $k_j\rightarrow k_j +1$, while $k_i+k_j$ is conserved
(as a result, $\alpha$ becomes a conserved quantity). 
In other words, $i$ is  the giver and $j$  is the taker,
while the probability of non-exchange is $1-R$.
One  simple way to give an advantage to an element with bigger $k$ is letting $R$  
dependent on $k_i$ and $k_j$ as below,
\begin{equation}
	R = \left\{
		\begin{array}{lc}
			1  &  (0 < k_i \leq k_j) \\
			\beta   & (k_i > k_j) \\
			0  & (k_i = 0) \\
		\end{array}\right.
\label{r}
\end{equation}
where $\beta$ is a constant in the range of $0\leq \beta  \leq  1$.
In this model, we can represent the system with three independent parameters: $N$, $\alpha$ and $\beta$.
When a distribution is given initially at $T = 0$, $P(k)$  will evolve as $T$ increases, 
and eventually reach a stationary distribution.
To express the cumulative distribution, we also define $P(\ge\!k)\equiv\Sigma_{k'=k}^\infty P(k')$ 

The parameter $\beta$ plays an important role in this model.
Two special cases of $\beta=0$ and $\beta=1$ have been previously discussed
in the context of conserved exchange processes \cite{ispolatov,dragulescu,chatterjee}.
The focus of this paper, however, is the general case of $0<\beta<1$.
When $\beta<1$, the time-reversal symmetry of the dynamics is broken, and at the same time the elements
with bigger $k$ (`the rich') get an advantage over those with smaller $k$ (`the poor').
Then this model becomes one of rich-get-richer mechanisms, which will
generate broad stationary distributions.
In the next section, we will show that the stationary distribution from this model exhibits the power law 
when $\beta$ is less than a certain critical value, and that this
critical value will depend on the value of $\alpha$.

\section{Stationary distributions}
First we look at the dynamics of an element with the quantity $k$ to focus 
on the evolution of an element. 
To gain one unit, an element should be chosen as the taker with the probability  $1/N$, and  
the probability of an element gaining one unit, $T_+(k)$, depends on the choice  
of the giver. Similarly the  probability of an element losing one unit, $T_-(k)$,
when chosen as a giver, can be found,
\begin{eqnarray}
T_+(k) &=& [1 - P(0) - (1-\beta) P(\ge\!k+1)],  \nonumber \\
T_-(k) &=& (1-\delta_{k0})[\beta+(1- \beta)P(\ge\!k)].
\label{P+P-}
\end{eqnarray}
Then, for an element, the expected change of $k$ after a time step, $\Delta k$, is
$[T_+(k)-T_-(k)]/N$.
Since $\Delta k$ is not proportional to $k$, the Gibrat's law is not satisfied. 
In a sense,  each element is performing the random walk if we regard $k$ as the position,  
while the transition probability found in Eq.\ (\ref{P+P-}) is  asymmetrical, position-dependent 
and time-varying.

If we use the continuous approximation as $N \rightarrow \infty$, the master equation can be obtained,
\begin{equation}
	\Delta P(k) = [P(k-1) T_+(k-1) - P(k) T_-(k) ] 
		- [P(k) T_+(k) - P(k+1) T_-(k+1)],
\label{DPk}
\end{equation}
where $\Delta P(k)$ is the expected change of $P(k)$ after one time step.
Then from the condition for the stationary distribution,
$\Delta P(k) = 0$ ($\forall k$, $k\ge0$),
we find that stationary distributions should satisfy these nonlinear equations,
\begin{equation}
	P(k+1)=\frac{T_+(k)}{T_-(k+1)} \, P(k)
		=\frac{1-P(0)-(1-\beta)P(\ge\!k+1)}{\beta+(1-\beta)P(\ge\!k+1)} \, P(k),
\label{Pk+1}
\end{equation}
for $k\ge0$, 
because, in Eq.\ ({\ref{DPk}), there are two parts, two terms each, and each part should be zero
when $\Delta P(k)=0$.
Even though we can theoretically find $P(k)$
as a function of $\alpha$ and $\beta$ using Eqs.\ (\ref{alpha}) and (\ref{Pk+1}),
these nonlinear equations are not easily solved analytically except for some special cases.

\subsection{Case of $\beta=0$}
This is a trivial winner-take-all situation. 
When $\beta = 0$, the rich will always win for every binary interaction.
Even without solving Eq.\ (\ref{Pk+1}), the stationary state and its distribution 
are trivially found.
In the stationary state, one element has all quantities, $k=\alpha N$, and the other
elements have no quantity, $k=0$; therefore the stationary distribution is
\begin{equation}
P(k) = \frac{N-1}{N}\delta_{k0}+\frac{1}{N}\delta_{k,\alpha N}.
\label{beta0}
\end{equation}
As $N\rightarrow\infty$, $P(k)$ becomes $\delta_{k0}$ approximately.

\subsection{Case of $\beta=1$}
When $\beta = 1$, the model describes the conserved random exchange process, which was already
discussed previously \cite{ispolatov,dragulescu}.
From Eq.\ (\ref{Pk+1}), we easily find  $P(k)$ exactly as $[1-P(0)]^kP(0)$.
After substituting $P(k)$ into Eq.\ (\ref{alpha}) to find $P(0)$, 
we obtain the stationary distribution,
\begin{equation}
	P(k)=\frac{1}{1+\alpha}\left(\frac{\alpha}{1+\alpha}\right)^k.
\label{beta1}
\end{equation}
As $\alpha\rightarrow\infty$, $P(k)$ becomes $(1/\alpha)\exp{[-k/\alpha]}$, 
which is the Boltzmann-Gibbs distribution.

\subsection{Case of $0<\beta<1$}
In this general case, the rich have an advantage over the poor, but lose to the poor from time to time.
This property keeps the stationary distribution balanced somewhere between those from two extreme cases 
discussed above.
Since the analytic method cannot be solely used in this case, the model is numerically investigated first.

After performing extensive numerical simulations while varying $N$, $\alpha$, and $\beta$,
we found that the stationary distributions exhibit the power law 
when $\alpha$ and $\beta$ satisfy a certain condition; 
that is, it is scaling when $(\alpha,\beta)$ is inside
a region in $(\alpha,\beta)$-space, represented by a condition such as $f(\alpha,\beta)<\epsilon$,
where $f(\alpha,\beta)$ and $\epsilon$ will later be found in this section.
In Fig.\ \ref{evolution}, we show a scaling case of $N =  10^5$, $\alpha = 1$, 
and $\beta = 0.1$.
\begin{figure}
\includegraphics[scale=0.3]{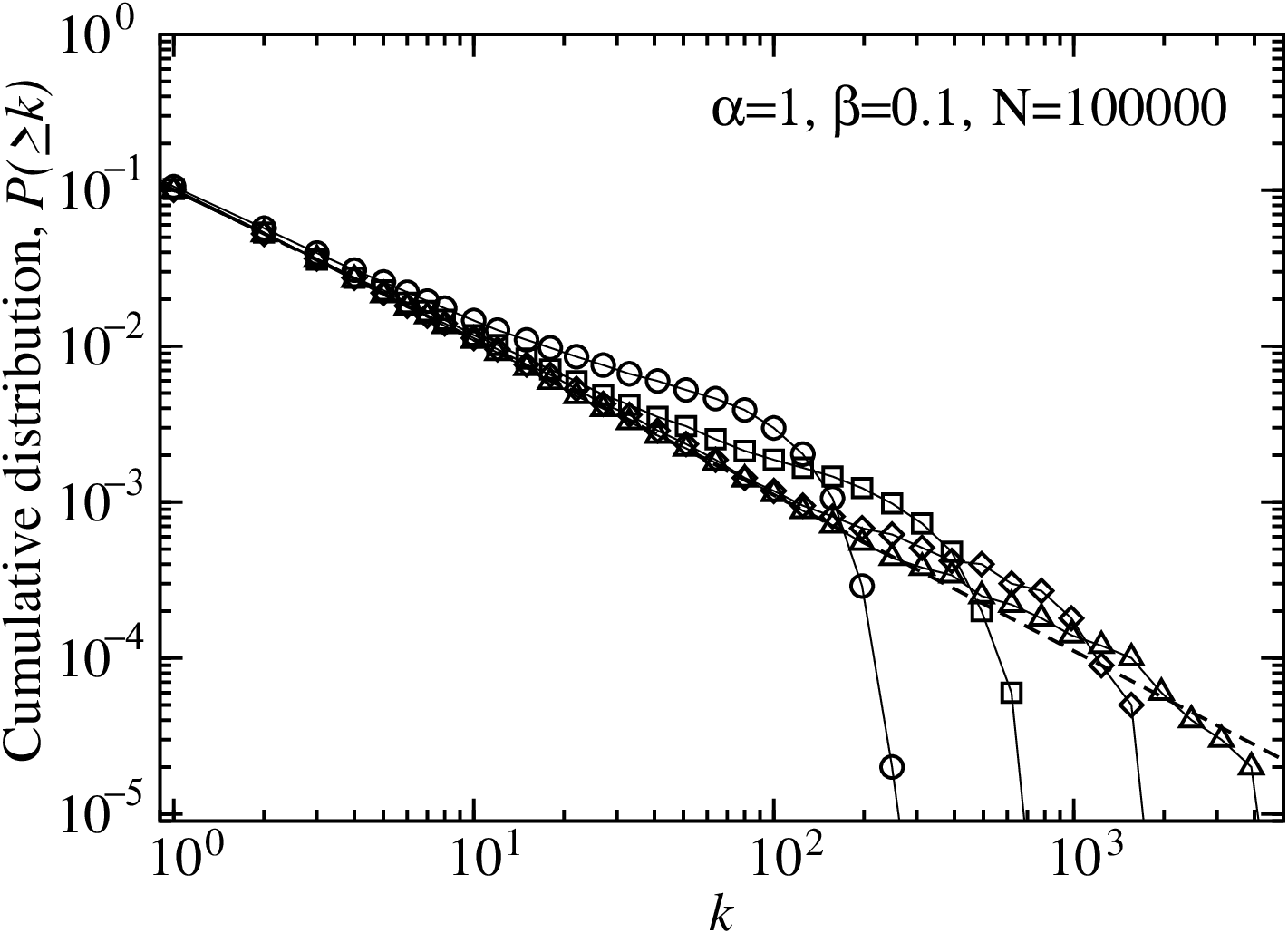}
\caption{\label{evolution}
Evolution of the cumulative distribution $P(\ge\! k)$
when $N = 10^5$, $\alpha = 1$, and $\beta = 0.1$ in a log-log plot.  
From an initial  distribution, $P(k) = \delta_{k1}$, 
we observe how $P(\ge\!k)$ evolves as the number of timesteps $T$ varies from 0 to 
$10^9$ ($\bigcirc$), $10^{10}$ ($\Box$), 
$10^{11}$ ($\Diamond$), $10^{12}$ ($\triangle$). 
We can observe that $P(\ge\!1)\simeq \beta(=0.1)$, which leads us to $P(0)=1-P(\ge\!1)\simeq 1-\beta$.
The dashed line represents the theoretical stationary distribution $P(\ge\!k)=1/(9k+1)$ at $\beta =  0.1$.
}
\end{figure}
As time increases, the initial distribution, $\delta_{k1}$, evolves to a power-law stationary
distribution, which is shown using the cumulative distribution, $P(\ge\!k)$.
(In  all simulations here, $\alpha$ is a positive integer, and the initial distributions of  the delta-function form, 
$\delta_{k\alpha}$, will be used.) 

One common property that stands out in all scaling cases like the one in Fig.\ \ref{evolution}
is that $P(0) \simeq 1-\beta$ whenever the distribution is scaling.
This property can be analytically proved by solving Eq.\ (\ref{Pk+1}) when $P(0)$ is given as $1-\beta$.
Since $P(\ge\!k+1)=1-\sum_{k'=0}^{k}\,P(k')$,
$P(k+1)$ can be represented as a function of $P(0)$, $\ldots$, $P(k)$, and $\beta$. 
When $k=0$, $P(1)$ can be found as a function of $P(0)$ and $\beta$,
and when $k=1$, $P(2)$ can also be found as a function of $P(0)$ and $\beta$ using $P(1)$ obtained already.
If we repeat this process, $\{P(k)|k\ge1\}$ will all be found as a function of $P(0)$ and $\beta$.
When we substitute $P(0)=1-\beta$, found numerically in scaling cases,
we can obtain $P(k)$ and $P(\ge\!k)$ in closed forms as below,
\begin{eqnarray}
P(k)  &=& \frac{\beta }{{1 - \beta }}\frac{1}{{[k + 1/(1 - \beta )][k + \beta /(1 - \beta )]}},\nonumber\\
P(\ge\!k)  &=& \frac{\beta }{{1 - \beta }}\frac{1}{{k + \beta /(1 - \beta )}}.
\label{PkECkE}
\end{eqnarray}
This is the Zipf's law, $P(k)\propto k^{-2}$ and $P(\ge\!k)\propto k^{-1}$, valid 
when $k(1-\beta)$ is  big enough.
The shape of $P(k)$ in Eq.\ (\ref{PkECkE}) does not depend on $\alpha$, 
but as we  will show later $\alpha$ will play a significant role in deciding 
whether the system  is scaling or not.

As shown above, the relation $P(0)=1-\beta$ is the condition for the scaling stationary distributions.
In other words, a  scaling condition for $\alpha$ and $\beta$ can be found if 
we find a condition with which the condition $P(0)=1-\beta$ holds.
To observe when the relation $P(0)=1-\beta$ holds, we find $P(0)$ for various $\alpha$ and $\beta$ values
using numerical simulations. 
In Fig.\ \ref{P0Efig}, we show $P(0)$ versus $\alpha$, and 
$P(0)$ versus $\beta$  when $N = 10^4$.
\begin{figure}
\includegraphics[scale=0.5]{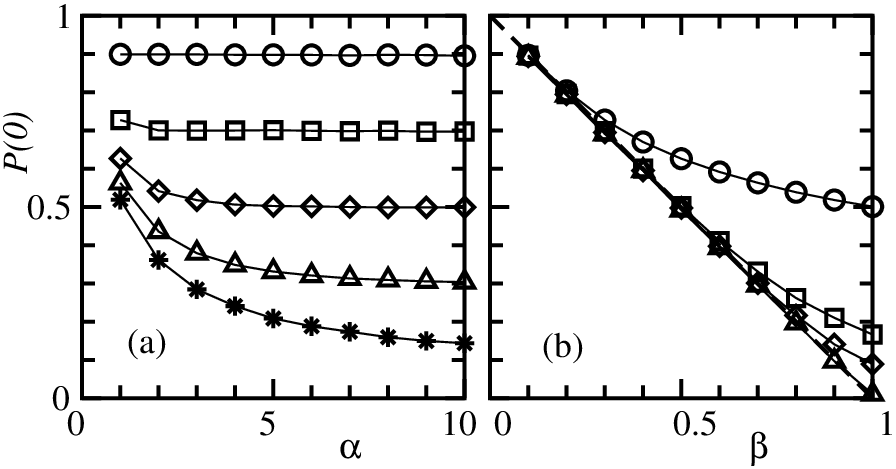}
\caption{\label{P0Efig}
When $N  = 10^4$ and $T = 10^9 \sim 10^{10}$, $P(0)$ values are 
found numerically for various $\alpha$ and $\beta$ values (averaged over 10 runs).
(a) $P(0)$  versus $\alpha$ when $\beta = 0.1$ ($\bigcirc$), 0.3 ($\Box$), 0.5 ($\Diamond$), 
0.7  ($\triangle$), 0.9 ($*$).
$P(0)$ is close to $1-\beta$ when $\alpha$ is greater than
a certain value for a given $\beta$.
(b) $P(0)$  when $\beta$ for $\alpha = 1$ ($\bigcirc$), 5 ($\Box$), 10 ($\Diamond$), 
100 ($\triangle$).  
$P(0)$ is close to $1-\beta$ when $\beta$ is less than
a certain value for a given $\alpha$.
Moreover, we observe that $P(0)=1/(1+\alpha)$
from Eq.\ (\ref{beta1}) are satisfied when $\beta=1$. 
The dashed line represents $P(0) = 1-\beta$.
}
\end{figure}
When $\beta$  is given, $P(0)$ equals to $1-\beta$ when $\alpha$ is greater than a certain critical value, $\alpha_c(\beta)$, 
and when $\alpha$ is given, $P(0)$ equals to $1-\beta$ when $\beta$ is less than a certain critical value, $\beta_c(\alpha)$
[$\beta=\beta_c(\alpha)$ is the inverse functions of $\alpha=\alpha_c(\beta)$]. 
Therefore we find that a critical relation exists for the system to  exhibit the scaling behavior,
and the boundary between the scaling region and non-scaling region is represented by $\alpha=\alpha_c(\beta)$.

How do we estimate this critical boundary in $(\alpha,\beta)$-space analytically? 
One possible  argument uses the highest $k$ value, $k_M$.
Since $N$ is finite in the model, the power law  will be valid only for a finite range of $k$, and the position of 
cutoff, $k_M$, depends on N, $\alpha$, and $\beta$. 
Especially when Eq.\ (\ref{PkECkE}) is satisfied (scaling cases), we can  estimate 
$k_M$ by solving the equation below, 
\begin{eqnarray}
	\alpha  &=& \mathop \Sigma \limits_{k = 0}^{k_M } kP(k) \nonumber \\ 
		&\simeq& 
			\frac{\beta }{{1 - \beta }}
			\int_0^{k_M } \!\! dk\left[ {\frac{k}{{[k + 1/(1 - \beta )][k + \beta /(1 - \beta )]}}} \right],
\label{alpha2} 
\end{eqnarray}
which is a modification of Eq.\ (\ref{alpha}) by letting $P(k) = 0$ when $k > k_M$  
(reasonable because $P(k)$ obtained from the continuous approximation is not valid
when $N$ is finite and $k$ is high). 
By solving Eq.\ (\ref{alpha2}), the estimated value  of $k_M$ for scaling cases is
\begin{equation}
	k_M  \simeq \frac{\beta }{{1 - \beta }}\,\beta ^{\frac{{ - 1}}{{1 - \beta }}}\, 
		\exp \left[ {\frac{\alpha }{{\beta /(1 - \beta )}}} \right].
\label{kM}
\end{equation}
Then we can find the ratio of the number of elements that are supposed to be in the 
region $k >  k_M$ to the total number of elements $N$, 
which can be obtained from $P(\ge\!k)$ at $k = k_M$, 
\begin{eqnarray}
	P(\ge\!k_M) 
		& \simeq & \frac{\beta }{{1 - \beta }}\frac{1}{{k_M }} \nonumber \\
		& \simeq & \beta ^{\frac{1}{{1 - 	\beta }}} \,
			\exp \left[ {\frac{{ - \alpha }}{{\beta /(1 - \beta )}}} \right] 
		 	\equiv  f(\alpha, \beta).
\label{CkME}
\end{eqnarray}

If the ratio, $f(\alpha,\beta)$, is small enough, these elements that were supposed to be in $k>k_M$ can
be regarded as additional elements with $k=0$, changing $P(0)$ into $P(0)+f(\alpha, \beta)$,
and they will not disrupt the stationary power-law distribution. 
However when $f(\alpha,\beta)$ is not small, the  whole distribution can be disrupted
[see how $P(0)$ affects all elements in Eq.\ (\ref{P+P-})], 
and the  distribution will settle into another type of stationary distributions, 
which  decay much faster than scaling ones do. 
Therefore, the scaling condition can be  written as $f (\alpha,\beta) < \epsilon$ where $0<\epsilon \ll 1$. 
In Fig.\ \ref{alphabeta}, we plot this condition when $\epsilon = 10^{-3}$, 
estimated from the simulation results.
\begin{figure}
\includegraphics[scale=0.6]{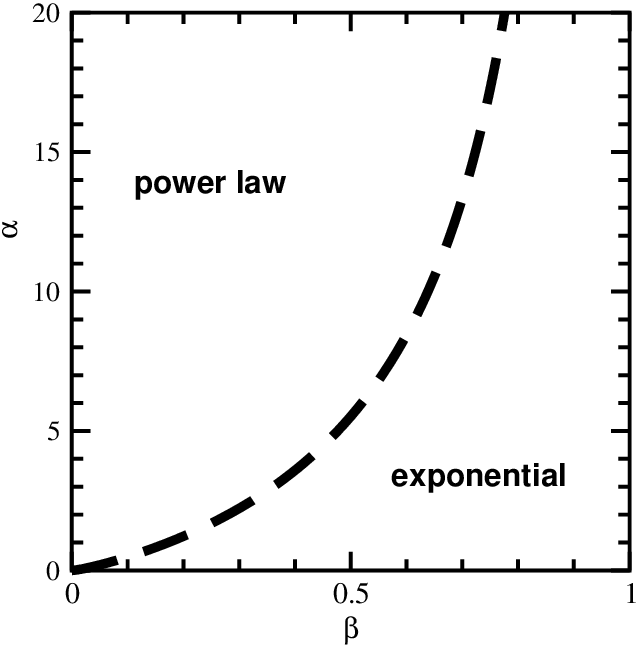}
\caption{\label{alphabeta}
The scaling condition in $(\alpha, \beta)$-space. The dashed line represents $f (\alpha, \beta) = \epsilon$ when
$\epsilon=10^{-3}$. As $\alpha$ increases, the range of $\beta$ for the power law approaches $0<\beta<1$.
}
\end{figure}
The critical boundary, $\alpha=\alpha_c(\beta)$, that separates the scaling region from the non-scaling region
was obtained from $f(\alpha,\beta)=\epsilon$,
\begin{equation}
	\alpha_c(\beta)=\frac{\beta}{1-\beta}\ln\!\left[\epsilon^{-1}\beta^{\frac{1}{1-\beta}}\right].
\label{alphac}
\end{equation}
The scaling region shown  in $(\alpha,\beta)$-space corresponds well with results in Fig.\ \ref{P0Efig}, 
and is  surprisingly big. 
If $\alpha$ is big enough, the system exhibits the power law for almost any value of $\beta$, 
which means that just a slight advantage given to the rich is  enough to make the system follow the Zipf's law. 
In Fig.\ \ref{examples}, we observe  several cases with various parameter values using numerical simulations. 
\begin{figure}
\includegraphics[scale=0.4]{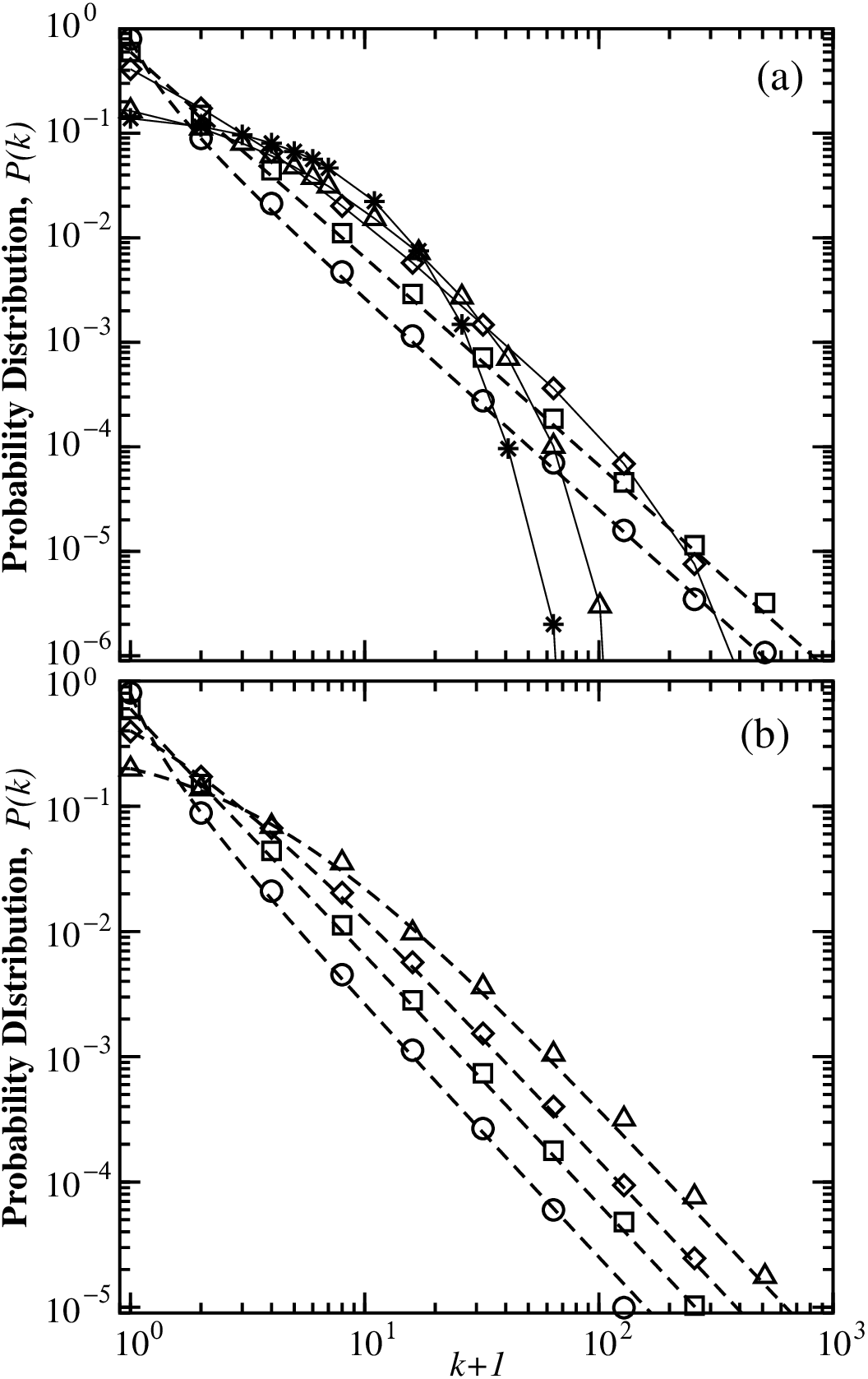}
\caption{\label{examples}
Stationary distributions for various $(\alpha, \beta)$ values. 
Dashed lines represent theoretical stationary probability distributions for given $\beta$ values, and
data points are  logarithmically binned for scaling cases. 
(a)  When $N = 10^6$ and $T = 7 \times 10^{11}$, $\alpha$  is fixed at 5, and $\beta = 0.2$ ($\bigcirc$),
0.4 ($\Box$), 0.6  ($\Diamond$), 0.8 ($\triangle$), 1.0 ($*$). 
The power-law distributions are observed clearly when $\beta = 0.2$  and 0.4. 
(b) Examples of various $(\alpha, \beta)$ values exhibiting power-law  distributions after $T = 10^{12}$: 
$N = 10^6$, $\alpha = 1$, $\beta = 0.2$ ($\bigcirc$); $N = 10^5$, $\alpha = 10$, $\beta  = 0.4$ ($\Box$);
$N = 10^4$, $\alpha = 100$, $\beta = 0.6$ ($\Diamond$); $N = 10^3$, $\alpha = 1000$, $\beta = 0.8$ ($\triangle$).
}
\end{figure}
In Fig.\ \ref{examples}(a), we fix $\alpha$ at  5 and vary $\beta$, to observe that $\beta_c(5) \sim 0.5$.
In Fig.\ \ref{examples}(b), we vary $\alpha$ and $\beta$ to observe that $\beta_c(\alpha)$ increases as
$\alpha$ increases.
When $(\alpha,\beta)$ is in the power-law region [$\beta<\beta_c(\alpha)$],
the shape of the stationary distribution is determined by $\beta$ only,
and $\alpha$ only changes the position of the cutoff, $k_M$.
On the other hand, the shape of the stationary distribution will be determined by both $\alpha$ and $\beta$
for non-scaling cases [$\beta>\beta_c(\alpha)$].
 
\section{Possible Applications}
So far we have proposed a simple model using general terms such as elements and  quantities. 
Here we discuss three examples where this mechanism can be applied.  

\subsection{Personal wealth}
The first example is the wealth distribution with people and their assets, which 
is known to exhibit the power law especially for the richest people.
In  a society, the population does not grow always, and their total amount of  assets can 
be assumed to be conserved. 
People also interact in many ways,  changing their assets, and the rich have an advantage over the poor. 
In our  model, $\alpha$ becomes the average amount of asset per person, and $\beta$ is the 
parameter representing the advantage for the rich.
Because $\alpha$ is usually big  enough, the power-law with Zipf exponent, $\gamma = 2$, 
will emerge for almost any  value of $\beta$, while empirical data shows $\gamma\simeq2.091$ \cite{newman}. 
There are other non-growth models for the power-law wealth distributions, 
which use the binary interactions of the traders \cite{ispolatov,chatterjee,xie}.

\subsection{City sizes}
The second example is the  distribution of city sizes with cities and their sizes. 
This is the  original Zipf's law, and the Zipf exponent has become famous for this  phenomenon 
(originally Zipf used the rank statistics and the exponent is 1,  which is equivalent to $\gamma = 2$
in our case). 
Our model can be applied to this case
when the number of cities is fixed, and the  overall population does not grow.
Here an interaction is the migration of a person (or a family)  from one city to another. 
People tend to move from a small city to a larger  city; 
hence, $\beta$ is the parameter representing this tendency. 
Then, the Zipf's  law will emerge from our model.
It will be unrealistic if $P(0)$ is not close to 0 because there is no empty cities usually.
But when $\alpha$ is big enough and $\beta$ 
is close to 1, the distribution will be still scaling and $P(0)$ will be close to 0.
Even with a drawback of
not taking account of the growth of cities from within unlike other models \cite{gabaix},
this mechanism has some merits to be regarded as another valid explanation of 
the Zipf's law: (1) the model produces the Zipf exponent naturally with a simple mechanism, 
(2) the migration of people between cities is well-represented by the model, 
(3) the attractiveness of the bigger cities is also well-represented.
There also can be a different approach. For example,
Zanette and Manrubia \cite{zanette} used the stochastic linear model, which
assumes neither the growth nor the binary interactions.

\subsection{Scale-free networks}
The last example is the network with nodes and their degrees. 
A network is  an entity that consists of nodes and links, 
while the degree of a node is the  number of links connected to a given node. 
In many systems represented by  networks, degrees of nodes have been found 
to follow power-law distributions  (hence called scale-free networks). 
Based on the mechanism of linear  preferential attachment proposed by Ref.\ \cite{barabasi}, 
many extended models have  been followed \cite{krapivsky,dorogovtsev,albert2,bollobas}. 
In these models, the assumption of growth of nodes and links is  crucial, 
and interactions between nodes are either ignored or used as an extra  feature \cite{albert2}. 
This approach is valid for many scale-free networks, but  not suitable for non-growing networks, 
in which node interactions are vital to  their dynamics. 
Our model can generate this kind of scale-free networks by  interpreting the interaction 
between nodes as the rewiring process. 
When nodes,  $i$ and $j$, are chosen, the rewiring process changes the link from $(i',i)$ to  $(i',j)$ 
where $i'$ is a {\em pivot} node chosen from nodes that are linked to $i$  
(loops and multiple links are allowed).  
Therefore, from our model, networks with power-law degree distributions can be  generated 
through only rewiring,
and the results will be presented in a forthcoming article \cite{p11}.
The network concept is actually related to  many scaling phenomena, 
since they can be represented by networks directly \cite{barabasi,albert,faloutsos,jeong,redner} 
or indirectly \cite{andersson}.

\section{Conclusions}
We have proposed a preferential-redistribution mechanism that generates  power-law distributions 
with the Zipf exponent for certain parameter ranges, and this scaling region in our parameter
space has been found analytically using some numerical results.
Since this scaling region is big enough and the mechanism is very simple, 
our model can be a good candidate to be used 
as a base mechanism for models describing some scaling phenomena, and
three possible applications have been discussed here. 
Like other models, our model has  limited applicability, but we believe that it can be extended to suit 
specific  needs as a part of more realistic models, or generalized to have more flexible  features.

\begin{acknowledgments}
This work was supported by grant No.\ R05-2002-000799-0 from the Basic 
Research Program of the Korea Science \& Engineering Foundation.
\end{acknowledgments}

\bibliography{revised2}

\end{document}